\documentclass[a4paper,superscriptaddress,showpacs,twocolumn]{revtex4}
\usepackage{epsfig}
\usepackage{graphics}
\usepackage{color}
\usepackage{amsmath}

\def\beq{\begin{equation}}
\def\eeq{\end{equation}}

\begin{document}

\title{Charged State of a Spherical Plasma in Vacuum}
\author{F.\,Cornolti}
\affiliation{Dipartimento di Fisica ``Enrico Fermi'' \& INFM , Universit\`a di
Pisa, Largo B. Pontecorvo 3, 56127 Pisa,
Italy }
\author{F.\,Ceccherini}
\email{ceccherini@df.unipi.it}
\affiliation{Dipartimento di Fisica ``Enrico Fermi'' \& INFM , Universit\`a di
Pisa, Largo B. Pontecorvo 3, 56127 Pisa,
Italy }
\author{S.\,Betti}
\affiliation{Dipartimento di Fisica ``Enrico Fermi'' \& INFM , Universit\`a di
Pisa, Largo B. Pontecorvo 3, 56127 Pisa,
Italy }
\author{F.\,Pegoraro}
\affiliation{Dipartimento di Fisica ``Enrico Fermi'' \& INFM , Universit\`a di
Pisa, Largo B. Pontecorvo 3, 56127 Pisa,
Italy }

\begin{abstract}
The stationary state of a spherically symmetric plasma configuration
is investigated in the limit of immobile ions and
weak collisions. Configurations with  small radii  are positively
charged  as a significant fraction of the electron
population evaporates during the equilibration process, leaving
behind an electron distribution function with an energy
cutoff.  Such charged plasma configurations are of interest for the
study  of Coulomb explosions and ion acceleration
from small clusters irradiated by ultraintense laser pulses  and for the investigation of ion bunches propagation in a plasma.
\end{abstract}

\maketitle
\section{Introduction}
The interaction of ultraintense laser pulses with solid targets leads
to the formation of plasmas with unusual
properties in terms of particle energy distributions and of spatial mass
and charge density  distributions. Such properties can be
exploited in order to obtain sources of high energy electromagnetic
radiation and  charged particle beams  with
unprecedented intensities and time and space resolutions.
For the intensities of present day ultrashort,  superintense laser
pulses, the energy that the ions in a target acquire due to direct 
interaction with the electromagnetic
fields of the  laser pulse is  usually small, while the  energy  of the  plasma
electrons can be of the order of tens of ${\rm KeV}$. These ``hot''
electrons expand until their ``pressure'' is
balanced by the electrostatic field that sets up due to spatial
charge separation \cite{allen1,allen2,allen3}. This process leads to a steady state configuration (SSC) which is reached in a time of the order of some electron plasma periods. Thus, since the ion response time
is much longer than that of the electrons, SSC can be achieved before the ions  can depart
significantly from their initial configuration. Afterwards ion acceleration takes place, as predicted theoretically 
\cite{allen1,allen2,allen3,nedelea,lontano,mora,ivanov,kovalev1,kovalev2,dorozhkina1,dorozhkina2,gurevitch, 
widner,betti},
and confirmed experimentally
\cite{hegelich,badziak,mackinnon1,mackinnon2,maksimchuk,clark1,clark2,krushelnick,hatchett,snavely, 
habara, mckenna}. 
Clearly, this description does not apply to highly relativistic regimes like those described in \cite{bulanov}.\\
The aim of this  paper is to present a combined analytical and
numerical investigation of three dimensional SSC characterized  by 
a hot electron plasma and a cold
(immobile) ion core. Such configurations are especially appealing because, contrary to one dimensional geometries, in three dimensional 
cases charged SSC are expected to set up. In fact, while in the former case an infinite energy is required in order to bring a charge to infinity, in the latter the energy necessary for electron evaporation is finite. In particular, we show that
the SSC charging up and the energy distribution of
the electrons  depend  on the ratio between the
radius of the ion core and the electron Debye length and on the
history of the electron expansion.
The understanding  of the SSC  charging up with immobile ions and of the electron energy
distribution is relevant  to  many experimental conditions as these
processes affect the way in which ions are
accelerated on longer time scales  when the constraint of a fixed ion
core is removed. In the case of the
Coulomb explosion of  a small cluster 
\cite{kumarappan1,kumarappan2,krainov,milchberg,ditmire1,ditmire2,ditmire3} these 
processes affect  the
value of  the maximum energy that the ions can acquire
in the acceleration process. Furthermore, in applications related to
proton imaging \cite{borghesi1,borghesi2,borghesi3} and to
the propagation of ion beams in  solid targets \cite{califano}, the
Coulomb repulsion and the screening effect  of the
electrons  can strongly affect the proton trajectories. This is also
the case for applications of proton laser
acceleration to  hadron therapy
\cite{bulanov1,bulanov2,esirkepov,orecchia1,brahme,weyrather,tsujii,orecchia2},
where a very precise collimation  of the proton beam  and a high
energy resolution are essential.
Moreover, as discussed in \cite{mulser},  the topic of the charging of
a spherical plasma in less extreme conditions can
play  a key role  in the study of dusty plasmas \cite{fortov}, and
aerosols \cite{kasparian}.\\
The paper is organized as follows. In Sec.\,\ref{SL} simple relationships are derived  analytically
on the basis of two schematic
models that are introduced in order to grasp the main features of
the plasma charging process in different
collisionality regimes. In Sec.\,\ref{NR} we present the results
of a series of  numerical  simulations obtained
with a one-dimensional Particle in cell code  (PIC) in  a spherical geometry
and with fixed ions. We then  compare the  numerically obtained 
charge values and electron  energy distributions  with those obtained  from the analytical models. Finally, conclusions are presented in Sec.\,\ref{conclusions}. \\
\section{Simplified models of the charging process}\label{SL}
In this section we discuss two simplified  models that allow us to
identify the main  physical features of  the
plasma charging process. These models rely on assigning a simplified
condition for the electrons to leave the ion core
and on two different rules for the energy redistribution of the
remaining electrons.\\
As a starting  configuration we assume the following electron and ion
density profiles
\begin{equation}\label{initial profile}
n_e(r) = n_i(r) = n_0 ~ \theta(1 - r/R),
\end{equation}
with $\theta(x)= 1$ for $x>0$ and  $\theta(x)= 0$ for $x<0$. Here
$R$ is the radius of the ion plasma core and $N_0 =
n_0 ( 4\pi R^3/3)$ the ion and electron initial particle number.
We denote by $N_e$ the time dependent number of electrons inside
the ion core. \\
Initially electrons have a Maxwellian energy distribution with
temperature $T_0$. Moreover, in these models the electron density
is taken  to be uniform inside the ion core. As a further simplification,
we assume that, on average, the radial crossing of the electron
trajectories   does not lead to a
relative redistribution of the charge in front and behind each
electron outside the core.  This simplification allows us
to assume that,  after leaving the ion core each electron moves as if
in an effective  time independent Coulomb field.
Hence, the  condition for an electron to reach infinity is that it
has  a positive total energy when it reaches the ion
core surface at $r=R$. In this model  the   particles which satisfy
this condition are said to ``evaporate'' and are
assumed to be lost  when they  reach $r=R$. On the contrary, the
electrons that have a negative energy at $r=R$ are
assumed to  remain inside the ion core.  Furthermore, we assume that
inside the ion core the electrons  move as free
particles.
The evaporation of the  electrons  with positive total energy at
$r=R$  changes the number of electrons $N_e$, the
total energy of the system inside the ion core and causes an energy
redistribution  of the remaining electrons.\\
We discuss the ``collisional'' regime and
  the ``collisionless'' one. In the first one,
  the electrons which are not evaporated
thermalize at a temperature $T$, which turns out to be a
decreasing function of time. In the second
regime no thermalization occurs, and the evaporation causes a
progressive depletion in the high energy tail of the
electron distribution function, which remains isotropic in velocity space.\\
In what follows, lengths will be measured in units of the initial
   Debye length $\lambda_d = ( T_0 / 4 \pi n_0 e^2 )^{1/2}$, with $e$ the
absolute value of the electronic charge, time $t$ in units of
$\omega_{pe}^{-1} =  (4 \pi e^2 n_0 / m_e)^{-1/2}$,
with $m_e$ the electron mass,
energies in units of the initial electron temperature
   $T_0$, velocities in units of the
initial electron thermal speed $v_{th,0} = \sqrt{T_0/ m_e}$,
mass in units of the electron mass and particle numbers in units
of $N_0$. Since inside the ion core the electron density is taken to be uniform, with the 
adopted normalization the normalized electron density $n_e$ and the normalized total 
number of electrons $N_e$ are numerically equal. \\

\subsection{Collisional regime \label{colregime}}
If the  electrons  inside the ion core are re-thermalized by
collisions, their velocity distribution function
   remains Maxwellian  and their time dependent  kinetic energy is
given by $U_{k} =3 N_e T/2 $.
   The electron evaporation rate is obtained by calculating the flux of
electrons with positive total energy through the
core surface.  Defining the time dependent quantity
$\phi_{T} = e \Phi_{R} / T =
(1-N_e) R^2 / 3 T$,  with $\Phi_{R}$ the
electrostatic potential at the
ion core surface, the positive total energy condition corresponds to
$v^2/ 2 T \geq \phi_{T}$. Thus  we obtain
\begin{equation} \label{thermal rate}
\frac{d N_e}{d t} = - \frac{3} {\sqrt{2 \pi} } ~ \frac{(1 + \phi_{T})}{\tau}
  ~ e^{-\phi_{T}} ~ N_e,
\end{equation}
where $\tau = R/\sqrt{T}$ is the
electron crossing time inside the ion core.
Analogously, the energy  flux $\Phi_U$ of the evaporating particle
can be computed
by noting that each evaporating electron carries away  the residual
energy $v^2 / 2 T - \phi_{T}$.
Then we obtain
\begin{equation} \label{energy flux}
\Phi_U =   \frac{3} {\sqrt{2 \pi} } ~ \frac{(2 + \phi_{T})}{\tau}
  ~ e^{-\phi_{T}} ~ N_e T.
\end{equation}
The total energy of the system can be written as  $U = {3}  N_e T/2 +
U_\Phi$  where  $U_\Phi$
is the electrostatic energy of the charged configuration
which increases as the electrons evaporate as
\begin{equation} \label{electroenergy}
\frac{d U_{\Phi}}{d t} = - \frac{2}{5} ~ R^2 (1-N_e)
\frac{d N_e}{d t}.
\end{equation}
Thus, from the total energy balance we obtain for the time change of
the kinetic energy of the system
$$
\frac{d (3 N_e T/2) }{d t}
= - \frac{3} {\sqrt{2 \pi} } \frac{(2 + \phi_{T})}{\tau}
~ e^{-\phi_{T}} ~ N_e T
$$
\begin{equation} \label{energy balance}
+ \frac{2}{5}R^2 (1-N_e) \frac{d N_e}{d t},
\end{equation}
which,  coupled to Eq. (\ref{thermal rate}), gives
the time evolution of the temperature $T$.\\
\subsection{Collisionless regime}\label{collisionless}
If on the contrary we assume that  plasma electrons inside the ion
core are not significantly affected by collisions, their
distribution function becomes non-Maxwellian. We assume that
the electron distribution  remains homogeneous in coordinate space
and isotropic in velocity space. Thus,
denoting by $N_{\cal E}$ the time dependent number of electrons
with kinetic energy (normalized on the initial temperatute $T_0$)
in the interval $[{\cal E}, {\cal E} + d {\cal E}]$, and introducing
the time dependent quantity
$\phi_0  = e \Phi_{R} = \frac{1}{3} (1-N_e) R^2$, which differs from 
$\phi_T$ in the previous
by the normalized temperature factor $1/T$, we obtain
\begin{equation} \label{energy decay}
\frac{d N_{\cal E}}{d t}
= - \frac{3}{2 \sqrt{2}} ~ \frac{\sqrt{{\cal E}}}{R} ~
\theta({\cal E} - \phi_0) ~ N_{\cal E}.
\end{equation}
This implies that the evaporation of
   the electron population with energy ${\cal E}$
stops at a well defined time $t = t_{\cal E}$,
where $t_{\cal E}$ is such that ${\cal E} = \phi_0 (t_{\cal E})$.
Therefore,
\begin{equation}\label{spectrum}
\begin{split} 
N_{\cal E} &= N_{\cal E}(0) e^{- t / t_d} \quad  {\rm for} \quad t \le t_{\cal E}\\
           &= N_{\cal E}(0) e^{- t_{\cal E} / t_d} \quad  {\rm for} \quad t > t_{\cal E}\\ 
\end{split}
\end{equation}
with $t_d = \left( 2 R \sqrt{2} / 3 \sqrt{{\cal E}} \right) ~  $
the ${\cal E}$-dependent decay time and
$N_{\cal E}(0)$ the electron kinetic energy distribution
at the initial time $t = 0$. We assume the electron velocities
at time $t = 0$ to be Maxwellian distributed,
hence the initial electron kinetic energy distribution  $N_{\cal 
E}(0)$ is given by
\begin{equation}\label{kinetic spectrum}
N_{\cal E}(0)
= N_{{\cal E},0} = \frac{2}{\sqrt{\pi}} e^{-{\cal E}} \sqrt{{\cal E}}.
\end{equation}
The electron number $N_e$, and therefore $\phi_0$, can thus be 
calculated performing
numerically, at fixed $t$, the integral
$
N_e = \int d {\cal E} N_{\cal E}.
$
Note that in this collisionless  model a rough estimate of the 
asymptotic electron number
could be obtained  by approximating the final electron distribution function
with the initial one for ${\cal E}<{\cal E}^*$, and with zero for 
${\cal E}>{\cal E}^*$.
We can then determine the cutoff energy ${\cal E}^*$  self consistently
by equating its value to  the electrostatic energy of the configuration
with charge
   $Q({\cal E}^*) =  \int_{{\cal E}^*}^\infty N_{{\cal E}}(0) d {{\cal E}}$,
\begin{equation} \label{th_formula}
{\cal E}^* = [Q({\cal E}^*) ~ R^2 / 3] .
\end{equation}
\subsection{Discussion of the analytical models}
Numerical integration of Eqs. (\ref{thermal rate},
\ref{kinetic spectrum}) shows that
both in the collisionless and in the collisional
regime the electron evaporation rate
is significantly reduced from its initial value when the quantities 
$\phi_T, \phi_0$ become of
order unity. This means that, in the collisional regime, the  evaporation
nearly stops when the electron electrostatic
energy, which is an increasing function of time,
is of the order of the electron temperature which decreases
with time. Whereas in the collisionless case the electron evaporation 
is significant
only until  the electron electrostatic energy is of the order of the
initial average electron kinetic energy.\\
The initial evaporation rate, which is obviously  the same  in both, the collisional
and in the collisionless case,  is obtained from either Eq.
(\ref{thermal rate}) or Eq. (\ref{energy decay}) which give
\begin{equation} \label{rate 0}
\left. \frac{d N_e}{d t} \right|_{t = 0} = -  \frac{3}{\sqrt{2 \pi}}  ~
\frac{1}{R}.
\end{equation}
 A linear approximation of $N_e$ is feasible as long as
\mbox{$~t \ll t^* ,  \tilde{t} ~$}  where
\begin{equation}\label{initevap time}
t^* =  \sqrt{2 \pi} R / 3 \end{equation}  is the initial characteristic evaporation time  and $\tilde{t}$ is defined  in 
the two different
collisionality regimes  by either  the condition $\left. \phi_T 
\right|_{\tilde{t}} \sim 1$ or
$\left. \phi_0
\right|_{\tilde{t}} \sim 1$  and  which can be roughly evaluated as
\begin{equation}\label{transition time}
\tilde{t} = {3} t^*/{R^2}.
\end{equation}
Therefore, if $t^* < \tilde{t}$, i.e., for
  ion core radii satisfying $R \leq \sqrt{3}$, the
  charging process continues until almost complete depletion
of the electron population.\\
Finally, we note that the time dependent electron energy distribution
predicted in the collisionless regime is highly non thermal.
The fact that the electron evaporation
only occurs for those particles with kinetic energy ${\cal E}$
satisfying ${\cal E} \geq \left. \phi_0 \right|_t$
causes a depletion of the high energy
tail of the electron distribution function,  as will be
  examined in detail in Sec. \ref{NR}.\\
\section{PIC simulations and comparison with the analytical results}\label{NR}
  Our PIC simulations are initialized with the
   electron and ion density profiles $n_e, n_i$  given by
Eq. (\ref{initial profile}). The initial electron distribution function
   is Maxwellian with temperature $T_0$.
At time $t = 0$ the electrons are allowed to move.
During their  expansion, the electrons
that reach the border of the simulation box with positive
total energy are removed.
%, as shown in Fig. ( (\ref{twoprofiles.eps})).
\begin{figure}
\resizebox{0.48\textwidth}{!}{%
   \includegraphics{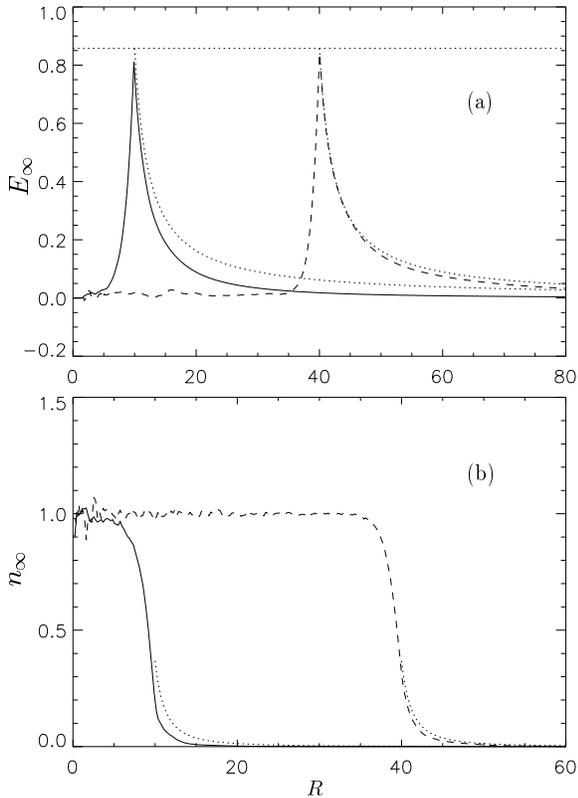}
}
\caption{Spatial profile of the electric field
$E_{\infty}$ (panel (a)) and electron density $n_{\infty}$ (panel (b))
for  $R=10$ (dashed line) and $R = 40$ (solid line).
For comparison they are plotted together with the profile  that would be 
obtained in the semi-infinite case (dotted lines) \cite{mora}. The straight line in panel (a) corresponds to the maximum value of the  electric field in the semi-infinite case, i.e, $E = \sqrt{2 / {\rm e}}$. }
\label{field_and_density.eps}
\end{figure}
Therefore, as the total number of electrons decreases with time,
the plasma acquires a net positive charge and an electrostatic potential
sets up. Finally a stationary state is reached where no more electrons reach the boundary.
We denote by
$N_{\infty}, n_{\infty}, E_{\infty},  N_{{\cal E},
\infty}$ the electron number, the electron density, the electric 
field profile, and the
electron kinetic energy distribution once SSC has been reached. 
As expected, our
simulations  show that these quantities depend on the ion core radius $R$.\\
The results regarding the profile of both the electric field 
$E_{\infty}$ and of the electron
density $n_{\infty}$ for two different  ion plasma radii $R$ are 
presented in Fig.\,\ref{field_and_density.eps}.
\begin{figure}
\resizebox{0.48\textwidth}{!}{%
   \includegraphics{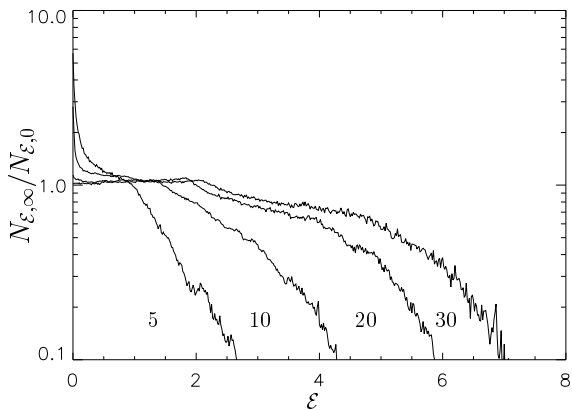}
}
\caption{Behaviour of the ratio between the final and the initial electron kinetic energy distributions for different ion plasma radii.}
\label{cut.eps}
\end{figure}
As one can see in Fig.\,\ref{field_and_density.eps}b, the electrons
which are outside the ion sphere are confined in a region whose width is of the order of a few Debye lengths.

\begin{figure}
\resizebox{0.48\textwidth}{!}{%
   \includegraphics{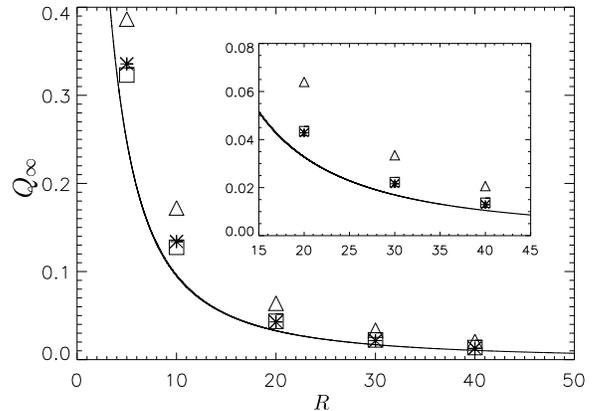}}
\caption{Comparison between the stationary state value
of the dimensionless charge inside the ion core
  predicted  by the collisional model (triangles),
   the collisionless  model (squares), the  PIC simulations (stars),
and the numerical solution of Eq. (\ref{th_formula}) (solid line).
A magnification
of the region $15 < R < 45$ is also shown in the figure. }
\label{Rfinale.eps}
\end{figure}

The numerical results regarding the electron kinetic energy distribution
in the stationary state are presented in Fig.\,\ref{cut.eps}. In the figure,
the ratio $N_{{\cal E},\infty} / N_{{\cal E},0}$,
is shown versus the electron kinetic
energy ${\cal E}$ in semi-logarithmic scale,
for several values of the ion plasma radius $R$.
These results show that the electron kinetic energy
distribution is highly non-thermal. It exhibits a cut in its high energy tail, 
and the cutoff energy increases with the ion plasma radius $R$. \\
Regarding the comparison with the 
semi-infinite, planar case,
  our results show
that both the electric field and the electron density are
  very similar to those presented in \cite{mora}
only as long as the ion core radius is greater  than several
tens of Debye lengths. Since, contrary to a one-dimensional configuration,  in  the
case of a three-dimensional configuration  the  energy required for the electrons to 
 evaporate is finite, the differences observed are mainly due to the cutoff in the  electron high energy tail. Such cutoff is responsible for the electron density depletion observed outside the ion core (see Fig.\,\ref{field_and_density.eps}b) and, consequently, for the corresponding electric field profile.
As shown in Fig.\,\ref{field_and_density.eps}a,
in the limit $R \gg 1$ the value of the dimensionless electric 
field at the surface of the ion core is almost independent of $R$, thus the net dimensionless charge confined within the ion core scales approximately as $1 / R $.\\
The comparison between the charge value $Q_{\infty} = 1 - N_{\infty}$
obtained numerically and that predicted
analytically is shown in Fig.\,\ref{Rfinale.eps}, for different values of the ion core radius $R$. Note that the value of $Q_{\infty}$ obtained numerically includes all SSC electrons, i.e., the electrons inside the ionic sphere and those in the surrounding halo.  
It is seen that the agreement among the numerical results and the values obtained in the collisionless regime is very good in the whole range $5-40$. With regard to the thermal model adopted for the collisional regime we remark that for small radii it predicts a moderately larger value of $Q_{\infty }$, but the two different regimes lead to very close values $Q_{\infty }$ in case of large radii. \\
Our results indicate that collisions can affect  the charging up
process only for small ion core  radii. This result can be
explained by noting that the potential due to the electron
expansion scales as $R^2$.  Thus, as the
potential barrier increases,  the fraction of the electron
population that, because of Coulomb collisions,  reaches a
positive total energy and can thus  leave the ion core
decreases. However, the collisional thermalization of the
distribution function assumed in Sec.\,\ref{colregime} can only be expected
to to apply  when the ion core radius is much larger
than the electron mean free path, whereas in most plasma
conditions the mean free path due to Coulomb collisions is much
greater than the Debye length. \\
\begin{figure}
\resizebox{0.48\textwidth}{!}{%
   \includegraphics{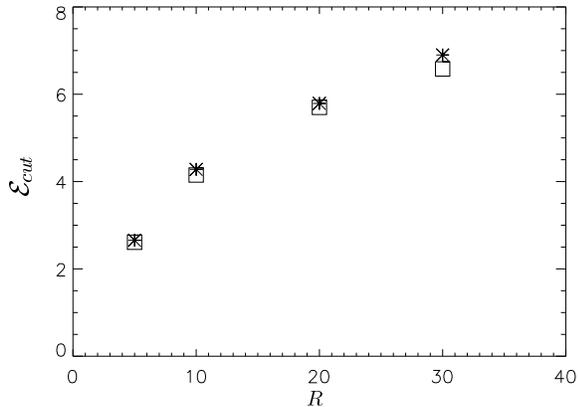}
}
\caption{Comparison between the cutoff energy
${\cal E}_{cut}$ as a function of the ion
core radius $R$ predicted by the collisionless model (squares) and
by the  PIC simulations (stars).}
\label{cut_comparison.eps}
\end{figure}
It is worth to note that a precise fit of the numerically-obtained plasma charge state  is given by the following Pad\'e approximation
\begin{equation}\label{Q}
Q_{\infty} \simeq \frac{1 + aR}{1 + bR + cR^2},
\end{equation}
with $a = 5.6\cdot 10^{-3}$, $b = 1.4\cdot 10^{-1}$ and $c = 5.5\cdot10^{-2}$. \\
With regard to the high tail of the electron kinetic energy distribution we define the cutoff energy ${\cal E}_{cut}$ as the energy satisfying the relation $ ({N_{{\cal E}, \infty}}/{N_{{\cal E}, 0}})|_{{\cal E}={\cal E}_{cut}}= 10^{-1}$. The analytical and numerical results obtained for  ${\cal E}_{cut}$  are shown  in Fig.\,\ref{cut_comparison.eps} for different values of the ion core radius $R$. Taking into account these results, we remark  that the collisionless model, although  not capable of reproducing the fine details of the whole electron kinetic energy distribution  observed in the simulations, nevertheless it predicts the cut in the  distribution high energy tail  with great accuracy.

\section{Conclusions}\label{conclusions}

In this paper we have investigated the charging up process of a 
spherically symmetric  plasma
configuration in vacuum in  the limit of immobile ions. Two different simplified models have been presented. With the help of these models we have established scaling laws relating the steady state total charge and electron energy distribution on the radius of the
ion core, normalized in terms of the initial electron Debye length. 
These scalings have been
validated by mean of spherical  one-dimensional PIC simulations. The agreement we find is overall very good. \\
Charged plasma configurations such as those investigated in this 
paper occur naturally  in the
interaction of ultraintense laser pulses with matter and are of 
interest, e.g., for setting the
initial conditions in the  study  of Coulomb explosions and ion 
acceleration from small clusters
irradiated by ultraintense laser pulses.\\
In particular, regarding the problem of cluster expansion the 
analytical and numerical results
that we have presented show that a spherically symmetric configuration
of cold ions and hot electrons, which is the typical starting 
configuration in cluster expansion
experiments, does not evolve towards a neutral configuration, in contrast with the  one-dimensional  planar case. This charging up effect can  strongly modify the maximum energy that the ions can gain and the typical timescale on which their  acceleration occurs.

\acknowledgments{
This work  was
supported by the INFM Parallel Computing
Initiative. Useful discussions with A. Macchi are gratefully acknowledged. }


\begin{thebibliography}{99}

\bibitem{allen1} 
J. E. Allen and J. G. Andrews,
 J. Plasma Physics, {\bf 4}, 187 (1970).
\bibitem{allen2} J. E. Crow, P. L. Auer and J. E. Allen,
  J. Plasma Physics, {\bf 14}, 65 (1975).
\bibitem{allen3} P. D. Prewett and J. E. Allen, 
J. Plasma Physics, {\bf 10}, 451 (1973).
\bibitem{nedelea} T. Nedelea and H. M. Urbassek,
 Phys. Rev. E {\bf 69}, 0546408 (2004).
\bibitem{lontano} M. Passoni and M. Lontano,
  Phys. Rev. E {\bf 69}, 026411 (2004).
\bibitem{mora} P. Mora, Phys. Rev. Lett. {\bf 90}, 185002 (2003).
\bibitem{ivanov} A. A. Ivanov and K.S. Serebrennikov,
 JETP Lett. {\bf 78}, 123 (2003).
\bibitem{kovalev1} V. 
F. Kovalev and V. Yu. Bychenkov, Phys. Rev. Lett. {\bf 90}, 185004 
(2003).
\bibitem{kovalev2} V. F. Kovalev, V. Yu. Bychenkov and V. T. Tikhounchuk,
 JETP Lett. {\bf 74}, 10 (2001).
\bibitem{dorozhkina1} D. S. Dorozhkina and V. E. Semenov,
 Phys. Rev. Lett. {\bf 81}, 2691 (1998).
\bibitem{dorozhkina2} D. S. Dorozhkina and V. E. Semenov,
 JETP Lett. {\bf 67}, 573 (1998).
\bibitem{gurevitch} A. V. Gurevitch, L. V. Pariskaya and L.P. Pitaievskii,
 J. Plasma Physics, {\bf 14}, 65 (1975).
\bibitem{widner} M. Widner, I. Alexeff and W. D. Jones,
 Phys. Fluids  {\bf 14}, 795 (1971).
\bibitem{betti} S. Betti, F. Ceccherini, F. Cornolti and F. Pegoraro,
 submitted for publication. Available at 
 {\tt http://xxx.lanl.gov/abs/physics/0405030}.
%plasma expansion, experiments
\bibitem{hegelich} M. Hegelich {\it et al.},
 Phys. Rev. Lett. {\bf 89}, 085002 (2002).
\bibitem{badziak} J. Badziak  {\it et al.},
Phys. Rev. Lett. {\bf 87}, 215001 (2001).
\bibitem{mackinnon1} A. J. Mackinnon {\it et al.},
Phys. Rev. Lett. {\bf 86}, 1769 (2001).
\bibitem{mackinnon2} A. J. Mackinnon {\it et al.},
 Phys. Rev. Lett. {\bf 88}, 215006 (2002).
\bibitem{maksimchuk} A. Maksimchuk and V. Yu Bychenkov,
 Phys. Rev. Lett. {\bf 84}, 4108 (2000).
\bibitem{clark1} E. L. Clark {\it et al.},
Phys. Rev. Lett. {\bf 84}, 670 (2000).
\bibitem{clark2} E. L. Clark {\it et al.},
Phys. Rev. Lett. {\bf 85}, 1654 (2000).
\bibitem{krushelnick} K. Krushelnick, 
Phys. Plasmas {\bf 7}, 2055 (2000).
\bibitem{hatchett} S. P. Hatchett {\it et al.}, Phys. Plasmas 
{\bf 7}, 2076 (2000).
\bibitem{snavely} R. A. Snavely {\it et al.}, 
Phys. Rev. Lett. {\bf 85}, 2945 (2000).
\bibitem{habara} H. Habara {\it et al.}, 
Phys. Rev. E {\bf 69}, 036407 (2004).
\bibitem{mckenna} P. McKenna {\it et al.}, 
Phys. Rev. E {\bf 70}, 036405 (2004).
\bibitem{bulanov} T. Zh. Esirkepov {\it et al.}
Phys. Rev. Lett.  {\bf 92}, 175003 (2004). 
%clusters
\bibitem{kumarappan1} V. Kumarappan, M. Krishnamurthy and D. Mathur, 
Phys. Rev. A {\bf 66}, 033203 (2003).
\bibitem{kumarappan2} V. Kumarappan, M. Krishnamurthy and D. Mathur,  
Phys. Rev. Lett. {\bf 87}, 085005 (2001).
\bibitem{krainov}  V. P. Krainov, M.B Smirnov, Phys. Rep. {\bf 370},
 237 (2002).
\bibitem{milchberg}  H. M. Milchberg, S. J. McNaught and E. Parra, 
Phys. Rev. E {\bf 64}, 056402 (2001).
\bibitem{ditmire1}  J. Zweiback, T. Ditmire and M. D. Perry,  
Phys. Rev. A {\bf 59}, R3166 (1999).
\bibitem{ditmire2}  T. Ditmire {\it et al.},
Nature (London) {\bf 398}, 489 (1999).
\bibitem{ditmire3}  T. Ditmire {\it et al.},
Nature (London) {\bf 386}, 54 (1997).
\bibitem{ditmire4}  T. Ditmire, R. A. Smith, J. W. G. Tisch and M. H. R. Hutchinson,
Phys. Rev. Lett. {\bf 78}, 3121 (1997).
\bibitem{ditmire5}  T. Ditmire {\it et al.},  
Phys. Rev. A {\bf 53}, 3379 (1996).
\bibitem{shao} Y. L. Shao  {\it et al.},  
Phys. Rev. Lett. {\bf 77} 3343 (1996).
\bibitem{snyder} E. M. Snyder, S. A. Buzza and A. W. Castleman Jr., 
Phys. Rev. Lett. {\bf 77} 3347 (1996).
%proton imaging 
\bibitem{borghesi1} M. Borghesi {\it et al.},
 Phys. Rev. Lett. {\bf 92}, 055003 (2004).
\bibitem{borghesi2} M. Borghesi {\it et al.},
 Phys. Rev. Lett. {\bf 88}, 135002 (2004).
\bibitem{borghesi3} M. Borghesi {\it et al.},
Phys. Plasmas {\bf 9}, 2214 (2002).
\bibitem{califano} F. Califano, F. Pegoraro and S. V. Bulanov,
Phys. Rev. E {\bf 68}, 066406 (2003).
%hadrontheraphy
\bibitem{bulanov1} S.V. Bulanov {\it et al.} Phys. Lett. A {\bf 299}, 240 
(2002).  
\bibitem{bulanov2} S.V. Bulanov and V. S. Khoroshkov, 
Plasma. Phys. Rep. {\bf 28}, 453 (2002).
\bibitem{esirkepov} T. Zh. Esirkepov {\it et al.}, 
Phys. Rev Lett. {\bf 89}, 175003 (2002).
\bibitem{orecchia1} 
R. Orecchia {\it et al.},  Clinical Reviews in Oncology/Hemathology 
{\bf 51}, 81 (2004).
\bibitem{brahme} A. Brahme, Int. J. Radiat. 
Oncol. Biol. Phys. {\bf 58}, 603 (2004).
\bibitem{weyrather} W. K. 
Weyrather and J. Debus, Clinical Oncology {\bf 15}, s23 
(2003).
\bibitem{tsujii} H. Tsujii, Eur. J. Cancer {\bf 37}, s251 
(2004).
\bibitem{orecchia2} R. Orecchia {\it et al.}, Eur. J. Cancer 
{\bf 34}, 459 (1998).
\bibitem{amaldi} U. Amaldi, analysis {\bf 1}, 1 (2003).
%dusty plasmas, aereosols, etc...
\bibitem{mulser} M. Kanapathipillai {\it et 
al.}, Phys. of Plasmas, {\bf 11}, 3911 (2004).
\bibitem{fortov} V. E. 
Fortov {\it et al.}, New. J. of Phys. {\bf 5}, 102 
(2003).
\bibitem{kasparian} J. Kasparian {\it et al.}, Science {\bf 
301}, 61 (2003).
\end{thebibliography}
\end{document}